\documentclass[twocolumn,nofootinbib,preprintnumbers,superscriptaddress,amsmath,amssymb,PRL]{revtex4}
\usepackage{graphicx}
\usepackage{gensymb}
\usepackage{dcolumn}
\usepackage{float}
\usepackage{mathptmx, courier, pifont}
\usepackage[scaled=0.92]{helvet}
\usepackage[T1]{fontenc}
\usepackage{textcomp}
\usepackage{color}
\usepackage[normalem]{ulem}

\usepackage[dvipsnames]{xcolor}
\usepackage[colorlinks=true,urlcolor=Blue,linkcolor=Blue]{hyperref}
\usepackage[all]{hypcap}
\usepackage{makecell}

\begin{document}



\title{Direct measurement of the mass difference of $^{72}$As-$^{72}$Ge \\rules out $^{72}$As as a promising $\beta$-decay candidate to determine the neutrino mass}
\author{Z.~Ge}\thanks{Corresponding author. Email address: zhuang.z.ge@jyu.fi}
\affiliation{Department of Physics, University of Jyv\"askyl\"a, P.O. Box 35, FI-40014, Jyv\"askyl\"a, Finland}%
\author{T.~Eronen}\thanks{Corresponding author. Email address: tommi.eronen@jyu.fi}
\affiliation{Department of Physics, University of Jyv\"askyl\"a, P.O. Box 35, FI-40014, Jyv\"askyl\"a, Finland}%
\author{A.~de Roubin}
\affiliation{Centre d'Etudes Nucl\'eaires de Bordeaux Gradignan, UMR 5797 CNRS/IN2P3 - Universit\'e de Bordeaux, 19 Chemin du Solarium, CS 10120, F-33175 Gradignan Cedex, France}%
\author{D.~A.~Nesterenko}
\affiliation{Department of Physics, University of Jyv\"askyl\"a, P.O. Box 35, FI-40014, Jyv\"askyl\"a, Finland}%
\author{M.~Hukkanen}
\affiliation{Department of Physics, University of Jyv\"askyl\"a, P.O. Box 35, FI-40014, Jyv\"askyl\"a, Finland}%
\affiliation{Centre d'Etudes Nucl\'eaires de Bordeaux Gradignan, UMR 5797 CNRS/IN2P3 - Universit\'e de Bordeaux, 19 Chemin du Solarium, CS 10120, F-33175 Gradignan Cedex, France}
\author{O.~Beliuskina}
\affiliation{Department of Physics, University of Jyv\"askyl\"a, P.O. Box 35, FI-40014, Jyv\"askyl\"a, Finland}%
\author{R.~de~Groote}
\affiliation{Department of Physics, University of Jyv\"askyl\"a, P.O. Box 35, FI-40014, Jyv\"askyl\"a, Finland}%
\author{S.~Geldhof}\thanks{Present address: KU Leuven, Instituut voor Kern- en Stralingsfysica, B-3001 Leuven, Belgium}
\affiliation{Department of Physics, University of Jyv\"askyl\"a, P.O. Box 35, FI-40014, Jyv\"askyl\"a, Finland}%
\author{W.~Gins}
\affiliation{Department of Physics, University of Jyv\"askyl\"a, P.O. Box 35, FI-40014, Jyv\"askyl\"a, Finland}%
\author{A.~Kankainen}
\affiliation{Department of Physics, University of Jyv\"askyl\"a, P.O. Box 35, FI-40014, Jyv\"askyl\"a, Finland}%
\author{\'A.~Koszor\'us}
\affiliation{Department of Physics, University of Liverpool, Liverpool, L69 7ZE,  United Kingdom}%
\author{J.~Kotila}
\affiliation{Finnish Institute for Educational Research, University of Jyv\"askyl\"a, P.O. Box 35, FI-40014, Jyv\"askyl\"a, Finland}%
\affiliation{Center for Theoretical Physics, Sloane Physics Laboratory Yale University, New Haven, Connecticut 06520-8120, USA}
\author{J.~Kostensalo}
\affiliation{Department of Physics, University of Jyv\"askyl\"a, P.O. Box 35, FI-40014, Jyv\"askyl\"a, Finland}%
\author{I.~D.~Moore}
\affiliation{Department of Physics, University of Jyv\"askyl\"a, P.O. Box 35, FI-40014, Jyv\"askyl\"a, Finland}%
\author{A.~Raggio}
\affiliation{Department of Physics, University of Jyv\"askyl\"a, P.O. Box 35, FI-40014, Jyv\"askyl\"a, Finland}%
\author{S.~Rinta-Antila}
\affiliation{Department of Physics, University of Jyv\"askyl\"a, P.O. Box 35, FI-40014, Jyv\"askyl\"a, Finland}%
\author{J.~Suhonen}
\affiliation{Department of Physics, University of Jyv\"askyl\"a, P.O. Box 35, FI-40014, Jyv\"askyl\"a, Finland}%
\author{V.~Virtanen}
\affiliation{Department of Physics, University of Jyv\"askyl\"a, P.O. Box 35, FI-40014, Jyv\"askyl\"a, Finland}%
\author{A.~P.~Weaver}
\affiliation{School of Computing, Engineering and Mathematics, University of Brighton, Brighton BN2 4JG, United Kingdom}%
\author{A.~Zadvornaya}
\affiliation{Department of Physics, University of Jyv\"askyl\"a, P.O. Box 35, FI-40014, Jyv\"askyl\"a, Finland}%
\author{A.~Jokinen} 
\affiliation{Department of Physics, University of Jyv\"askyl\"a, P.O. Box 35, FI-40014, Jyv\"askyl\"a, Finland}%

%
%
\date{\today}
\begin{abstract}
We report the first direct determination of the ground-state to ground-state electron-capture $Q$-value for the $^{72}$As to $^{72}$Ge decay by measuring their atomic mass difference
utilizing the double Penning trap mass spectrometer, JYFLTRAP. The $Q$-value  was measured to be 4343.596(75)~keV,
which is more than a 50-fold improvement in precision compared to the value in the most recent Atomic Mass Evaluation 2020. Furthermore, the new $Q$-value was found to be 12.4(40)~keV (3.1 $\sigma$) lower. With the significant reduction of the uncertainty of the ground-state to ground-state $Q$-value value combined with the level scheme of $^{72}$Ge from $\gamma$-ray spectroscopy, we confirm that the five potential ultra-low $Q$-value ${\beta^{+}}$-decay or electron capture transitions are energetically forbidden, thus precluding all the transitions as possible candidates for the electron neutrino mass determination. However, the discovery of small negative $Q$-values opens up the possibility to use $^{72}$As for the study of virtual $\beta$-$\gamma$ transitions.

\end{abstract}
\maketitle
\section{Introduction}

Neutrino oscillation experiments have given indirect evidence for finite neutrino masses through the observation of neutrino mixing.
The fact that neutrinos are massive is the strongest demonstration that the Standard Model (SM) of electroweak interactions is incomplete and new physics beyond the SM must exist. The oscillation experiments cannot answer the question of the possible Majorana nature of neutrinos and their absolute mass scale~\cite{Suhonen1998,Avignone2008,Ejiri2019}. Present techniques, which guarantee a model-independent approach for direct measurements of the electron (anti)neutrino  mass, are based on a kinematical analysis near the endpoint of $\beta$-decay spectra~\cite{Drexlin2013,Aker2019,Gastaldo2014,Gastaldo2017,Alpert2015,Faverzani2016,Croce2016}.  So far, the study of the tritium $\beta$-decay end-point by means of the electrostatic spectrometer KATRIN (KArlsruhe TRitium Neutrino) has yielded the most stringent upper limit for the electron antineutrino mass, 1.1 eV (90\% Confidence Level (C. L.))~\cite{Drexlin2013,Aker2019}. The  allowed $\beta^{-}$-decay transition of $^{3}$H(1/2$^{+}$)$\rightarrow$ $^{3}$He(1/2$^{+}$), with a $Q$-value of 18.59201(7) keV~\cite{Myers2015}, is employed in the KATRIN experiment.

For $\beta$ decays, the fraction of decay events that fall into an energy interval just below the end-point energy ($Q_{\beta }$) is  proportional to $Q_{\beta }^{-3}$;  while for electron capture (EC), the event rate proportionality is even steeper.
%
%
This implies that isotopes with the lowest $Q$-value are desirable~\cite{Ferri2015}. $^{187}$Re has been considered for its low $Q$-value of 2.492(30)$_{\rm stat}$(15)$_{\rm syst}$  keV~\cite{Basunia2017,Nesterenko2014} and because it can be used as a bolometric detector~\cite{Nucciotti2012,Ferri2015}. MARE (Microcalorimeter Arrays for a Rhenium Experiment) is the corresponding long-term project carrying the expectations for the future direct neutrino mass measurements~\cite{Nucciotti2012,Ferri2015}. Originally based on the development of fast $^{187}$Re bolometers, MARE is now also including $^{163}$Ho for its low $Q$-value of 2.833(30)$_{\rm stat}$(15)$_{\rm sys}$ ~\cite{Eliseev2015,Ranitzsch2017},  suggested  as a unique opportunity for a self-calibrated and high statistics experiment exploiting the enhancement in sensitivity due to the closeness of the $^{163}$Ho EC $Q$-value and the $^{163}$Dy atomic $M$-lines.    

The search for other isotopes that could undergo a low (especially ultra-low (<1 keV)) $Q$-value $\beta$-decay or EC is of great interest for possible future (anti)neutrino mass determination experiments~\cite{Mustonen2010,Mustonen2011,Suhonen2014}.
These transitions include decays into excited states in the daughter nucleus. Whether these type of decays are energetically possible or not, can be experimentally determined by measuring the ground-state to ground-state decay $Q$-value and the excitation energy of the state in the daughter nucleus.

The ultra-low $Q$-value decay  branch of $^{115}$In (9/2$^{+}$) to the first excited state of $^{115}$Sn (9/2$^{+}$) was first revealed by Cattadori \emph{et al.}~\cite{Cattadori2005}. The $Q$-value of this branch was independently measured to be  0.155(24) keV~\cite{Mount2009} by Penning trap measurements at Florida State University and 0.35(17) keV~\cite{Wieslander2009} at the University of Jyv\"askyl\"a.

\begin{figure}[htbp]
\centerline{}
	\includegraphics[width=1.0\columnwidth]{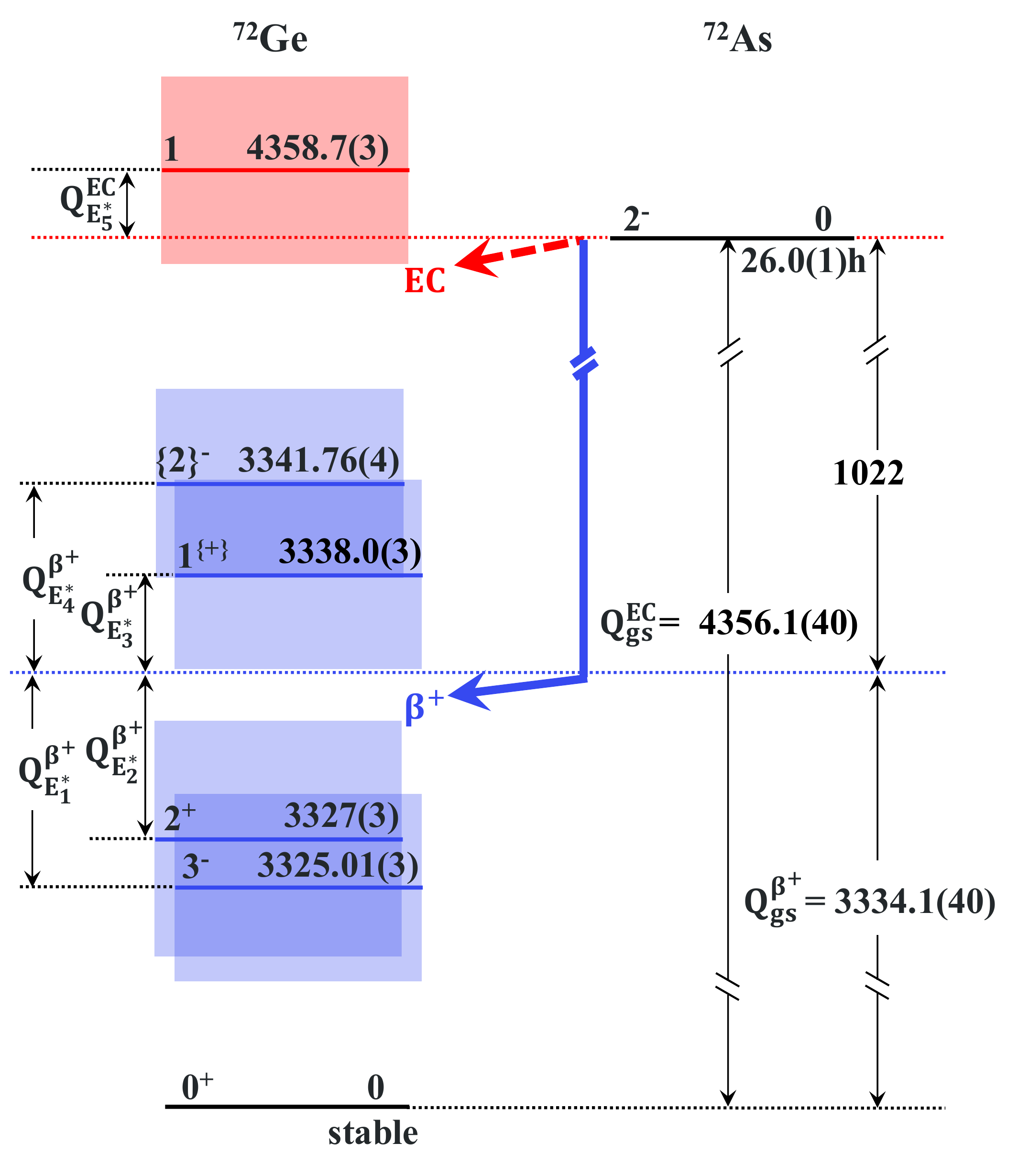}
	\caption{(color online) Partial $\beta^+$/EC decay scheme of $^{72}$As ground state to excited states in $^{72}$Ge. $Q_\textrm{gs}$ for both $\beta^+$ and EC decays are from AME2020~\cite{Huang2021,Wang2021} and the energies of the excited states in $^{72}$Ge from~\cite{NNDC}. The colored shaded bands show the $1\sigma$ uncertainty in the $Q$-value, which is almost solely defined by the uncertainty in $Q_{gs}$. The states below the thick dashed red line marked ``EC'' and blue line marked ``$\beta^+$'' are energetically possible with electron capture and $\beta^+$ decay, respectively. See also Table~\ref{table:low-Q1}.
	}
   \label{fig:level-scheme}
   \centerline{}
\end{figure}

The ground-state to ground-state decay $Q$-value ($Q_{gs}$) for $\beta^-$ and EC decays is the difference of atomic masses of the decay parent ($M_p$) and the decay daughter ($M_d$):
\begin{equation}
   Q_{gs}^{\beta^{-}} = Q_{gs}^{EC}  = (M_p - M_d)c^2    
\end{equation}
and for ${\beta^{+}}$-decay
\begin{equation}
Q_{gs}^{\beta^{+}} = Q_{gs}^{EC} - 2m_ec^2,    
\end{equation}
where $m_e$ is the rest mass of the electron and $c$ is the speed of light in vacuum. Combining the  $Q_{gs}$ values with the excitation energy of the decay daughter state, $E{_i}{^*}$ yields  the  ground-state to excited-state  $Q$-value:
\begin{equation}
Q_{E{_i}{^*}}^{\beta^{\pm}/EC} = Q_{gs}^{\beta^{\pm}/EC}  -  E{_i}{^*}.
\end{equation}
For the EC decay, the atomic binding energy $B_j$, where $j$ denotes the atomic shell of the captured electron, needs to be taken into account.
Electron binding energies are typically known to high precision.

The $Q$-values of the decays to excited states in the daughter are normally known with only 1 keV precision or worse~\cite{Gamage2019,Wang2021}. The excitation energies $E_i^*$ in the daughter nuclei are typically known with sub-keV precision~\cite{NNDC} while the $Q_{gs}$-values are poorly known and commonly lack a value from a direct measurement. Presently, Penning-trap mass spectrometry (PTMS) is the most precise and accurate method for determining atomic masses and $Q_{gs}$-values and routinely reaches sub-keV precision. 

Several potential ultra-low $Q$-value $\beta$-decay candidates have recently been studied via PTMS, for example $^{89}$Sr, $^{131}$Cs, $^{135}$Cs and  $^{139}$Ba~\cite{DeRoubin2020,Sandler2019,Karthein2019a}. Among them, one promising candidate of the first-forbidden unique $\beta^{-}$-decay transition, $^{135}$Cs (7/2$^{+}$, ground state)$\rightarrow ^{135}$Ba (11/2$^{-}$, second excited state),  was confirmed to be an ultra-low $Q$-value  decay channel. A $Q$-value of 0.44(31) keV was determined via a measurement at the JYFLTRAP Penning trap~\cite{DeRoubin2020}. 

\begin{table}[!htb]
   \caption{Final states in $^{72}$Ge after EC/$\beta^+$ decay of $^{72}$As 2$^{-}$ ground state that potentially have a low $Q$-value. The $Q_\textrm{gs}$-value is from AME2020~\cite{Wang2021} and the excitation energies from~\cite{NNDC}. Spin-parity assignments enclosed by braces indicates that these are uncertain, which results in an uncertainty in the decay type, indicated by a  \{?\}  in the fourth column. 1st FNU represents 1st forbidden non-unique. For the ground state decay, the  $Q_{gs}^{EC}$-value is given. See also Fig.~\ref{fig:level-scheme} and text for details.} 
  \begin{ruledtabular}
   \begin{tabular*}{\textwidth}{ccccc}
State, $i$ & $E^*$ (keV)  &  $J^\pi$ &Decay type &    $Q$ (keV)  \\
\hline\noalign{\smallskip}
1&3325.01(3)&   3$^{-}$&  {$\beta^{+}$:  Allowed}    &  8.9(40)     \\ 
2&3327(3)&   2$^{+}$&     { $\beta^{+}$: 1st FNU}      &  6.9(50)    \\
3&3338.0(3)&   1$^{\{{+}\}}$&  {  $\beta^{+}$:   1st FNU\{?\} }  &  -4.1(40)     \\
4&3341.76(4)&   \{2\}$^{-}$& {  $\beta^{+}$:  Allowed\{?\} }  &  -7.9(40)     \\
5&4358.7(3)&   1& {  EC:  Allowed\{?\} }  &  -2.8(40)    \\ \hline
gs&0 & 0$^{+}$& & 4356(4)   \\
   \end{tabular*}
   \label{table:low-Q1}
   \end{ruledtabular}
\end{table}

In this work, we report on the $Q_{gs}$-value determination performed with the JYFLTRAP Penning trap for the promising candidate nucleus $^{72}$As ($t_{1/2} = 26.0(1)$~h). The decay daughter, $^{72}$Ge, has five closely-lying excited states where the EC or $\beta^+$ decay of $^{72}$As could proceed with an ultra-low $Q$-value as shown in Fig.~\ref{fig:level-scheme}. The candidate transitions are listed in Table~\ref{table:low-Q1}. Three out of the five transitions are possibly allowed and the other two are first-forbidden non-unique (FNU) transitions.  It is worth noting that the allowed transitions are particularly interesting cases since they have larger branching ratios, enabling the accumulation of more data in a shorter time period and potentially making the case more lucrative  for direct (anti)neutrino mass determination~\cite{Suhonen2014,Gamage2019}. In addition, because the decay transition is driven by a single decay matrix element, the beta spectral shape is universal, which makes the analysis of the $\beta$-decay spectrum simpler.

The ground-state to ground-state $Q_{gs}^{\beta^{+}/EC}$ value in AME2020 originates from ${\beta^{+}}$-decay measurements of $^{72}$As(${\beta^{+}}$)$^{72}$Ge which fully defines the $Q$-value~\cite{1950Me55,NSR1968VI05}. As many previous Penning-trap experiments have already demonstrated large deviations from the $Q_{gs}$-values and masses deduced from $\beta$-decay or reaction experiments~\cite{Hardy1977,Eliseev2011a,Nesterenko2019}, direct mass measurements are strongly desired. With the current precision of the $Q_{gs}$-value, it is difficult to distinguish which of these transitions are energetically possible and further, actually fall into the ultra-low $Q$-value (< 1~keV) category. This puzzle is now solved for $^{72}$As by a direct mass difference measurement of $^{72}$As-$^{72}$Ge.

\section{Experimental description}
\begin{figure}[!htb]
    \centerline{}
   \includegraphics[width=0.99\columnwidth]{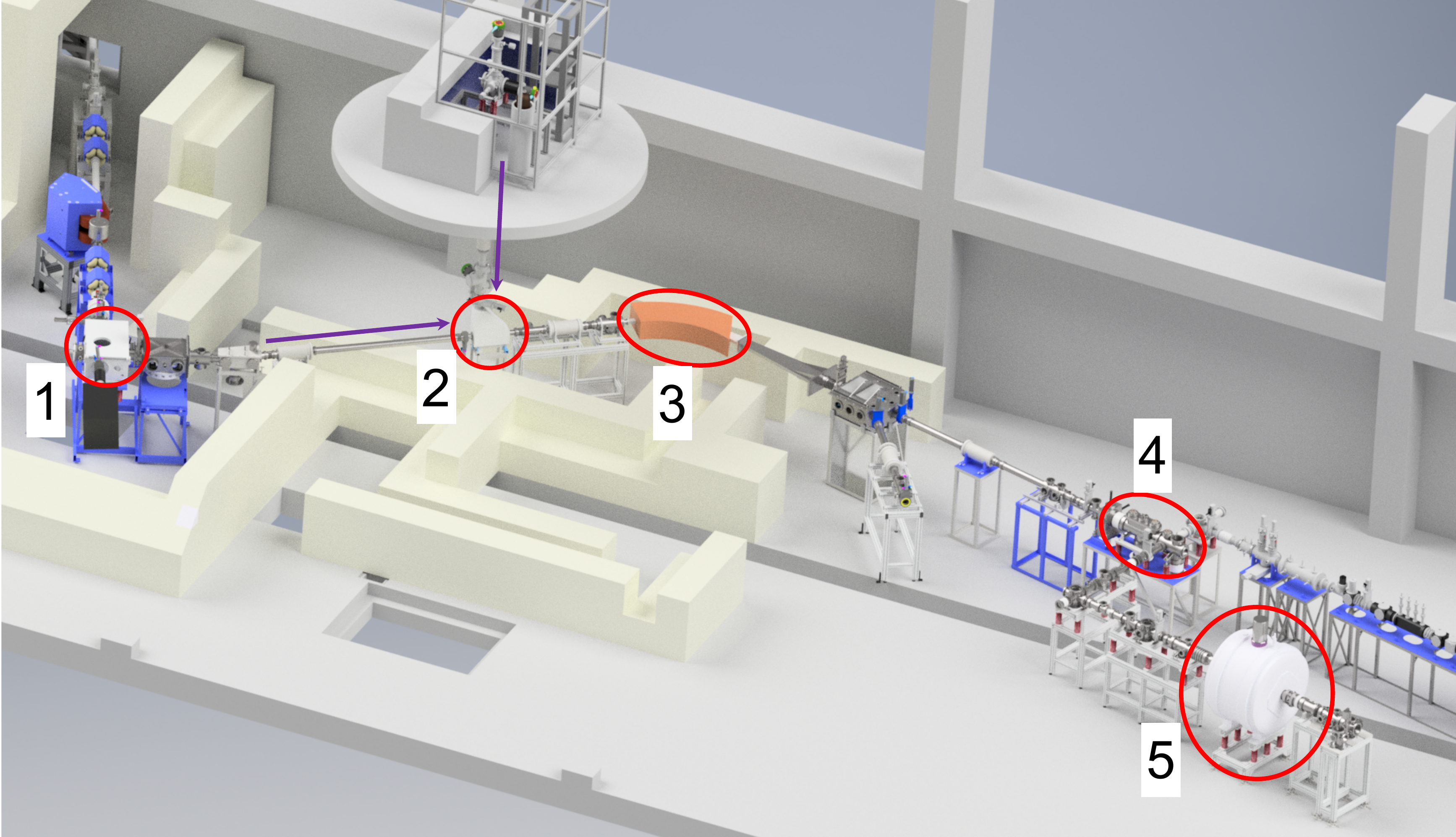}
   \caption{(Color online)   Schematic overview of the IGISOL facility. The  $^{72}$As$^{+}$ and $^{72}$Ge$^{+}$ ions were produced with deuteron-induced  fusion reactions at the IGISOL target  chamber (1). After production and extraction from the gas cell, the ions were guided through an electrostatic bender (2),  mass number ($A=72$) selected  with a dipole magnet (3), cooled and bunched in the RFQ cooler-buncher (4) and finally used for the mass-difference measurement in the JYFLTRAP Penning trap setup (5). }
   \label{fig:igisol}
\end{figure}

The $Q_{gs}$-value measurement was carried out at the Ion Guide Isotope Separator On-Line facility (IGISOL) using the JYFLTRAP double Penning trap mass spectrometer~\cite{Eronen2012}, located at the University of Jyv\"askyl\"a~\cite{Moore2013,Kolhinen2013}. See Fig.~\ref{fig:igisol} for the layout of the facility.
To produce the ions of interest at IGISOL, a thin germanium target with a thickness of about 2~mg/cm$^2$ was bombarded with a 9-MeV deuteron beam from the K-130 cyclotron. The reaction simultaneously produced both $^{72}$As$^+$ and $^{72}$Ge$^+$ ions.

The  produced ions are stopped in the gas cell of the IGISOL light-ion ion guide~\cite{Huikari2004} through collisions with high-purity helium gas at a pressure of about 100 mbar. During this process, the highly charged ions recombine to become predominantly singly charged. The recoils exit the gas cell through a small nozzle into a sextupole ion guide (SPIG) ~\cite{Karvonen2008}. The ions are guided via the SPIG into high vacuum and get accelerated to 30 keV of energy. 
A magnetic dipole mass separator with a mass resolving power of about 500 is sufficient to reject all but $A = 72$ ions, where A is the mass number. After the separation, the mass number selected ions are transported through an electrostatic switchyard housing a fast kicker electrode used to chop the beam to have an optimum number of ions. After the switchyard the ions are injected into a radiofrequency quadrupole (RFQ) cooler-buncher~\cite{Nieminen2001}, which is used to cool and bunch the beam. Finally, the bunches are transported to the downstream JYFLTRAP double Penning trap for the actual frequency ratio measurement.

The JYFLTRAP double Penning trap consists of two cylindrical traps located in a 7~T superconducting  magnet.  The cooled and bunched  ions are confined in a homogeneous magnetic field and a quadrupolar electrostatic potential inside the traps. The first trap, the purification trap, is used as a high-resolution mass separator, while the second trap, precision trap, is utilized for a high-precision mass determination by employment of the conventional time-of-flight ion-cyclotron-resonance (TOF-ICR) method~\cite{Koenig1995,Graeff1980}, or the application of the phase-imaging ion-cyclotron-resonance (PI-ICR) technique~\cite{Nesterenko2018,Eliseev2014,Eliseev2013}. 

The ion beam contained $^{72}$Ga$^+$ as a co-produced impurity.
In the first  trap an isobarically pure sample of ions was prepared by the mass-selective buffer gas cooling method~\cite{Savard1991}, which provides a typical resolving power $M/\Delta M \approx 10^{5}$.
 To prepare a clean sample of $^{72}$Ge$^+$, it was enough to use the mass-selective buffer gas cooling method to remove $^{72}$Ga$^+$, $^{72}$As$^+$ and any other ion species present in the beam. 
Preparation of a clean sample of $^{72}$As$^+$ required an additional cleaning step. The $^{72}$Ge$^+$ ions were removed with the buffer gas cooling method but to remove $^{72}$Ga$^+$ ions, a higher resolution Ramsey cleaning method was required~\cite{Eronen2008a}.

The $Q_{gs}$ value determination is based on the measurement of the cyclotron frequency
 \begin{equation}
\label{eq:nuc}
\nu_{c}=\frac{1}{2\pi}\frac{q}{m}B
\end{equation} 
for both the decay parent and decay daughter ions. Here $q/m$ is the charge-to-mass ratio of the stored ion and $B$ the magnetic field strength. The $Q_{gs}$-value is obtained through the cyclotron frequency ratio
\begin{equation}
    R = \frac{\nu_{c,Ge}}{\nu_{c,As}},
\end{equation}
where $\nu_{c,As}$ is the cyclotron frequency for $^{72}$As$^+$ and $\nu_{c,Ge}$ for $^{72}$Ge$^+$.
During this experiment, we  alternated between $^{72}$As$^+$ and $^{72}$Ge$^+$ cyclotron frequency measurements every few minutes to minimize contribution of the magnetic field fluctuation in the measured cyclotron frequency ratio. Still, 
a linear interpolation was used to obtain the magnetic field at the moment of the parent cyclotron frequency measurement.

In this work, the PI-ICR technique was used to measure the cyclotron frequencies~\cite{Nesterenko2018,Eliseev2014,Eliseev2013}. This technique is about 25 times faster reaching a certain precision compared to the TOF-ICR method.
In particular, the measurement scheme number 2 described in~\cite{Eliseev2014} was utilized to directly measure the cyclotron frequency.
Two timing patterns, one called ``magnetron'' and the other ``cyclotron''  were used, see ~\cite{Nesterenko2018}. These patterns are otherwise identical except for the switching on instant of the $\pi$-pulse that converts the ions' cyclotron motion to magnetron. In the ``magnetron'' pattern the ions predominantly revolve in the trap for a time duration $t$ (accumulation time) with magnetron motion while in the ``cyclotron'' pattern the ions revolve with cyclotron motion. The exact knowledge of the switch-on time difference $t$ is essential. The used patterns produce so-called magnetron and cyclotron spots or phases on the position-sensitive micro-channel plate (MCP) detector~\cite{PS-MCP}.
%
Additionally, it is necessary to measure the motional center spot. With these data, it is then possible to obtain the angle between the cyclotron and magnetron motion phases with respect to the center spot
\begin{equation}
\label{eq:alphac}
   \alpha_c = \alpha_+-\alpha_-,
\end{equation}
where $\alpha_+$ and $\alpha_-$ are the polar angles of cyclotron and magnetron phases, respectively. Finally, the cyclotron frequency $\nu_{c}$  is deduced from
\begin{equation}
\label{eq:nuc2}
\nu_{c}=\frac{\alpha_{c}+2\pi n_{c}}{2\pi{t}}.
\end{equation}
The measurement is set up so that $\alpha_c$ will be small in order to minimize systematic shifts due to image distortion by choosing $t$ to be as close to integer-multiples of $\nu_c$ period as possible.
$n_{c}$ is the number of complete revolutions during the phase accumulation time $t$.

The accumulation time $t$ for both $^{72}$Ga$^+$ and $^{72}$As$^+$ ions during the interleaved measurement was chosen to be 321~ms in order to ensure that the cyclotron spot was not overlapping with any possible isobaric, isomeric or molecular contamination.
%
This accumulation time $t$ was also specified to the nearest integer-multiple of period of $\nu_c$ to minimize the angle $\alpha_c$. This ensured minimal influence from the interconversion of magnetron and cyclotron motions~\cite{Eliseev2014,Kretzschmar2012c} and also minimized the image distortion shift. In these measurements $\alpha_c$ did not exceed a few degrees.

\begin{figure}[!htb]
   \includegraphics[width=0.99\columnwidth]{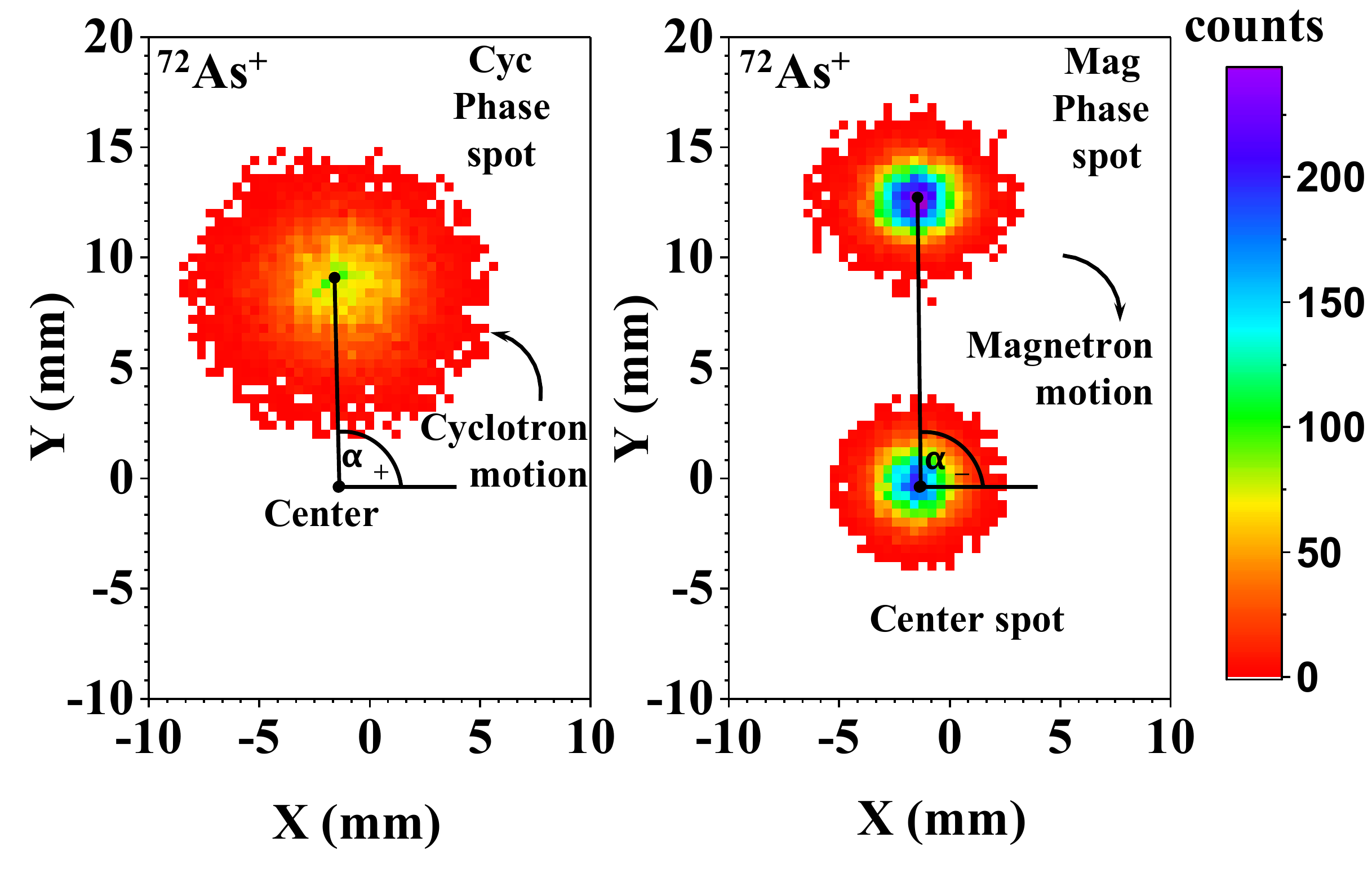}
   \caption{(Color online) Center, cyclotron phase and magnetron phase spots of $^{72}$As$^{+}$ on the position-sensitive MCP detector after the PI-ICR excitation pattern with an accumulation time of 321 ms. The figure comprises all data recorded in the experiment. The cyclotron phase spot is displayed on the left side and the magnetron phase spot with a center spot on the right. The angle difference between the two spots relative to the center spot is utilized to deduce the cyclotron frequency of the measured ion species. The color bar illustrates the number of ions in each pixel.}
   \label{fig:2-phases}
\end{figure}

The collected ``magnetron'' and ``cyclotron'' phase spots of $^{72}$As$^+$ ions are plotted in the left and right panels of Fig.~\ref{fig:2-phases}. The delay of the cyclotron motion excitation was repeatedly scanned over one magnetron period and the final extraction delay was varied over one cyclotron period to account for any residual magnetron and cyclotron motion that could shift the different spots. These constituted in a total of $5\times5=25$ scan points for both magnetron and cyclotron phase spots. As the decay parent  $^{72}$Ge$^+$ and the decay daughter $^{72}$As$^+$ ions were produced simultaneously at IGISOL, a direct doublet measurement was realized after separation and purification of the samples. The measurement of the $\nu_{c}$ of the ions $^{72}$Ge$^+$ and $^{72}$As$^+$  was performed continuously for a total duration of about 6.5 hours.%



The frequency measurement directly yields the $Q_{gs}^{EC}$-value (see Eq.~\ref{eq:Qec}) via the cyclotron frequency ratio $R$
 \begin{equation}
\label{eq:Qec}
Q_{gs}^{EC} = (R-1)(M_d - m_e)c^2+\Delta{B_{m,d}},
\end{equation}
where $M_d$ is the atomic mass of the decay daughter ($^{72}$Ge).
$\Delta B_{m,d}$ is a term that takes into account the electron binding energy difference of the decay parent-daughter atoms ($\approx$ 2.6 ~eV) using ionization energies from the National Institute of Standards and Technology (NIST)~\cite{NIST_ASD}.
Since both the parent and daughter have the same $A/q$, the mass-dependent error effectively becomes negligible compared to the statistical uncertainty achieved in the measurement. Additionally, due to the fact that the mass difference $\Delta M/M$ of $^{72}$Ga and $^{72}$As is smaller than $10^{-4}$, the contribution to the $Q_{gs}^{EC}$-value from the mass uncertainty (0.08 keV/c$^2$) of the reference $^{72}$Ge is negligible.


\section{Results and discussion}
The data was collected by initiating a $\nu_c$ measurement of $^{72}$As$^{+}$ for four full scan rounds followed by measurement of $^{72}$Ge$^{+}$ also for four full scan rounds. After, a center spot was recorded with $^{72}$Ge$^+$ ions. In total, these steps lasted about seven minutes in total and were repeated over a period of 6.5 hours.

For each repetition, positions of each spot were determined using the maximum likelihood method and the phase angles were calculated to deduce the cyclotron frequencies. 
Consecutive fitted cyclotron frequencies of $^{72}$Ge$^{+}$ were linearly interpolated to the time of the measurement of $^{72}$As$^{+}$. This interpolated frequency was used to deduce the cyclotron resonance frequency ratio $R$. In this manner, a total of about 70 frequency ratios were obtained.
Contribution of  temporal fluctuations of the magnetic field  to the final frequency ratio uncertainty was less than 10$^{-10}$ since the frequency measurements of the ion pair were tightly interleaved.

The incident ion rate was limited to a maximum of 5 detected ions/bunch with the median value being around 2 ions/bunch. Bunches with more than 5 ions were rejected from the analysis in order to reduce a possible cyclotron frequency shift due to ion-ion interactions~\cite{Kellerbauer2003,Roux2013}. Countrate-class analysis~\cite{Kellerbauer2003} was used to confirm that the frequency was indeed not shifting.

\begin{figure}[!htb]
   \includegraphics[width=0.99\columnwidth]{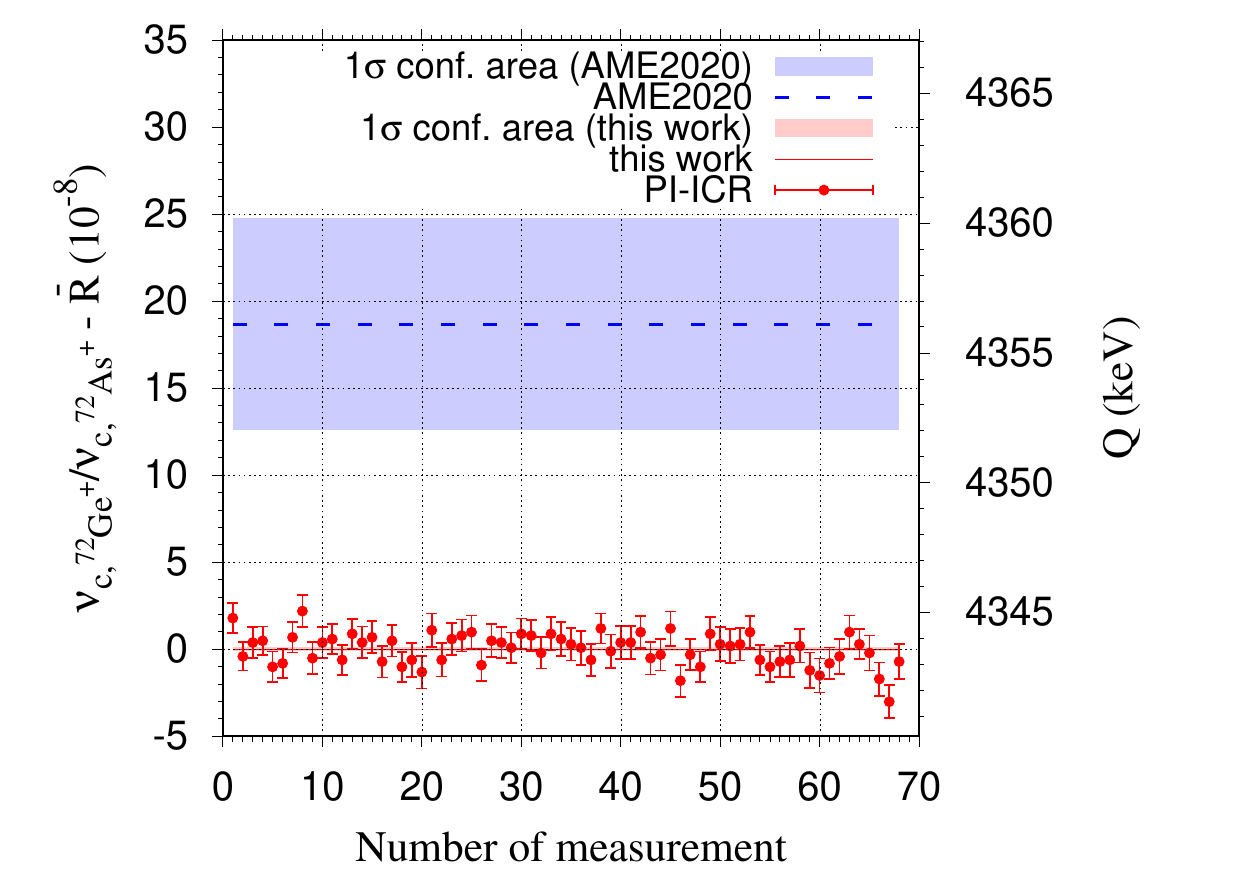}
   \caption{(Color online) Comparison of results from this work and AME2020. The left axis shows the corresponding frequency ratio deviation from the measured value and the right axis the $Q$-value. The red points are the data points and the solid horizontal red line with the shaded area the final value. The dashed blue line is the AME2020 value (shaded area is the $1\sigma$ uncertainty). 
   %
   }
   \label{fig:ratio}
\end{figure}

The frequency shifts due to ion image distortions were well below the statistical uncertainty. This was ensured by keeping $\alpha_c$ of Eq. (\ref{eq:alphac}) small ($< 4$ degrees). The weighted mean ratio $\overline{R}$ of the single ratios for PI-ICR data was calculated along with the inner and outer errors. The ratio of inner and outer errors, otherwise known as the Birge ratio~\cite{Birge1932}, was found to be 0.987.
%
%
The larger of the errors, the outer error, was taken as the final uncertainty. In Fig.~\ref{fig:ratio}, the results of the analysis are compared to the literature values. 

The final frequency ratio $\overline{R}$ and the resulting $Q_{gs}^{EC}$-value are $1.000 064 835 1(11)$  and $4343.596(75)$~keV, respectively.

To check the reliability of the above interpolation method, a polynomial fitting method~\cite{Nesterenko2018} was also used to deduce the frequency ratio.
The temporal evolution of the measured cyclotron frequencies $\nu_{c,p}$(t) for parent $^{72}$As$^{+}$ ions  and $\nu_{c,d}(t)$  for daughter $^{72}$Ge$^{+}$ ions can be described with  the same polynomial function $f(t)$ and the frequency ratio $R$:
\begin{equation}
\label{eq:poly}
\nu_{c,p}(t)=f(t),
\nu_{c,d}(t)=R\nu_{c,p}(t)=Rf(t).
\end{equation} 

\begin{figure}[!htb]
   \includegraphics[width=0.99\columnwidth]{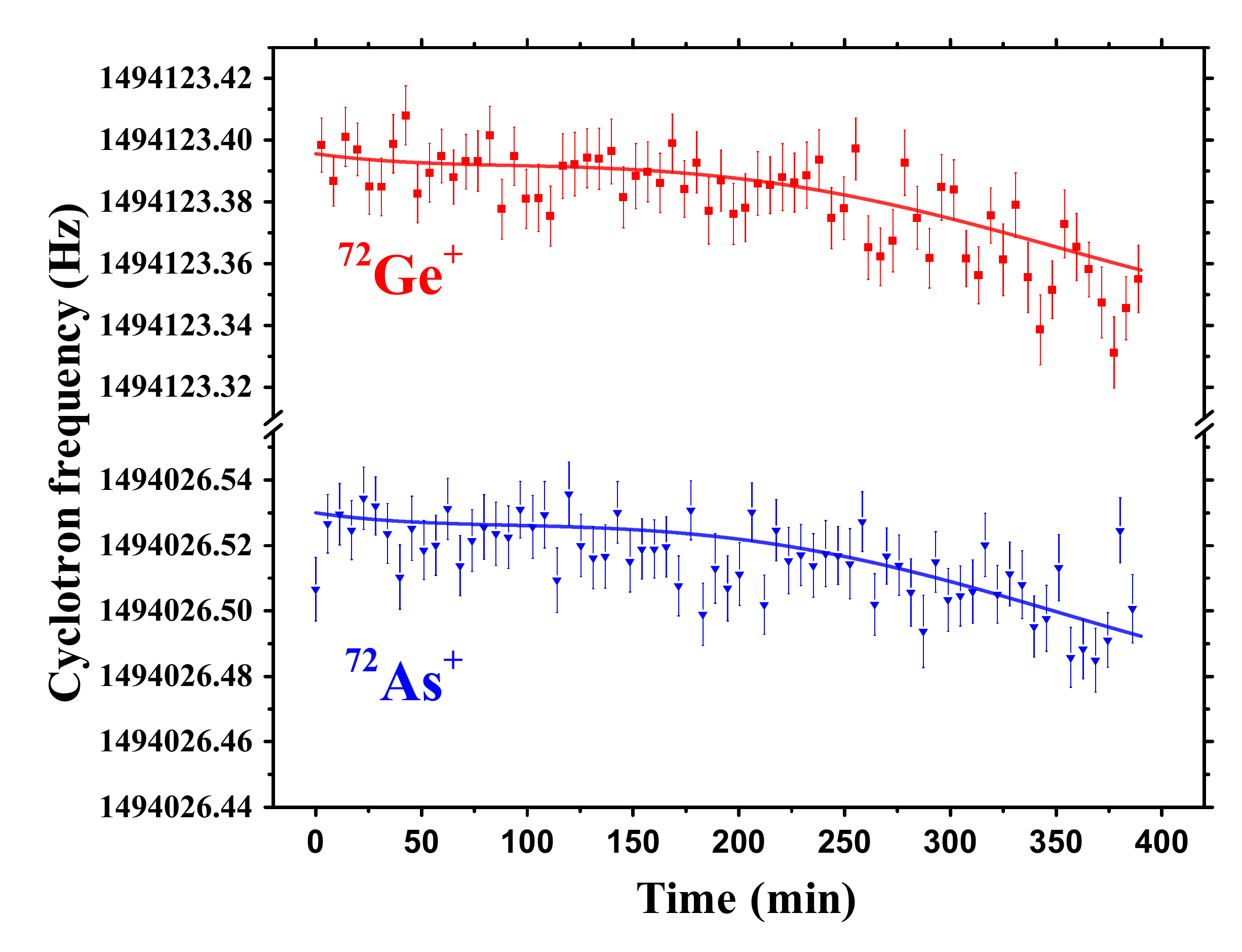}
   \caption{(Color online) Cyclotron frequency ratio determination using a simultaneous  polynomial fit to the measured  cyclotron frequency data. The reduced ${\chi}^2$ was 0.95. See text for more details.}
   \label{fig:poly}
\end{figure}

The order of the polynomial was chosen to be four, which gives the smallest reduced $\chi^2$ of the fit.
The result is shown in Fig.~\ref{fig:poly}, where individual frequency points are shown with the fit. 
The frequency ratio obtained from the polynomial fit agrees well with the one obtained from the linear interpolation analysis within a combined 1$\sigma$ uncertainty.


\begin{table}[!htb]
\caption{Final results based on the analysis of the mean cyclotron frequency ratio between the daughter ($^{72}$As$^{+}$) and parent ($^{72}$As$^{+}$) ions.  $Q_{gs}^{EC}$-value and the mass-excess (ME) of $^{72}$As determined in this work in comparison to the AME2020 values~\cite{Huang2021}.}
\begin{ruledtabular}
   \begin{tabular*}{\textwidth}{ccc}
& $Q_{gs}^{EC}$ (keV)& ME (keV/$c^2$) \\
\hline\noalign{\smallskip}
This Work&   4343.596(75)&  -68242.305(106) \\
AME2020 &     4356(4)&  -68230(4) \\
   \end{tabular*}
   \label{table:Qvalue}
\end{ruledtabular}
\end{table}

\begin{table}[!htb]
   \caption{$Q$-values for the decay candidates to the excited states of the daughter nucleus $^{72}$Ge obtained in this work compared to the values derived using AME2020 $Q_{gs}^{EC}$~\cite{Huang2021,Wang2021}. All data in the table are in keV. The first column gives the experimental excitation energy $E^{*}$~\cite{NNDC} of the daughter state in $^{72}$Ge. The second and third columns give the $Q$-value using the $Q_{gs}$ from AME2020 and from this work, respectively. The last column shows the confidence ($\sigma$) of the $Q_{gs}$ being negative.}
  \begin{ruledtabular}
   \begin{tabular*}{\textwidth}{cccc}
E$^{*}$& \makecell[c]{$Q$-value \\ (AME2020)} &\makecell[c] {$Q$-value \\(This work)}& \makecell[c]{$Q/\delta Q$ \\(This work)}  \\
\hline\noalign{\smallskip}
 3325.01(3)   &8.9(40)  &-3.42(8) & 43 \\ 
 3327(3)   &6.9(50)  &-5.4(30) &  1.8 \\
 3338.0(3) &-4.1(40)  &-16.41(31)  & 53 \\
 3341.76(4)  &-7.9(40)  &-20.17(8) &  238 \\
 4358.7(3) &-2.8(40)  &-15.11(31)   & 49 \\
   \end{tabular*}
   \label{table:low-Q}
   \end{ruledtabular}
\end{table}

The final $Q_{gs}^{EC}$-value and the mass-excess of $^{72}$As obtained from the mean cyclotron frequency ratio is given in  Table~\ref{table:Qvalue}. The  $Q_{EC}$-value from this work is a factor of 54 more precise and 12.4(40)~keV smaller than the value in AME2020~\cite{Huang2021}. 
The mass-excess value of $^{72}$As was improved by a factor of 38. It has an additional 0.08~keV/c$^2$ uncertainty contribution from the uncertainty of the reference daughter $^{72}$Ge ion mass.

\begin{figure}[!htb]
   \includegraphics[width=0.9\columnwidth]{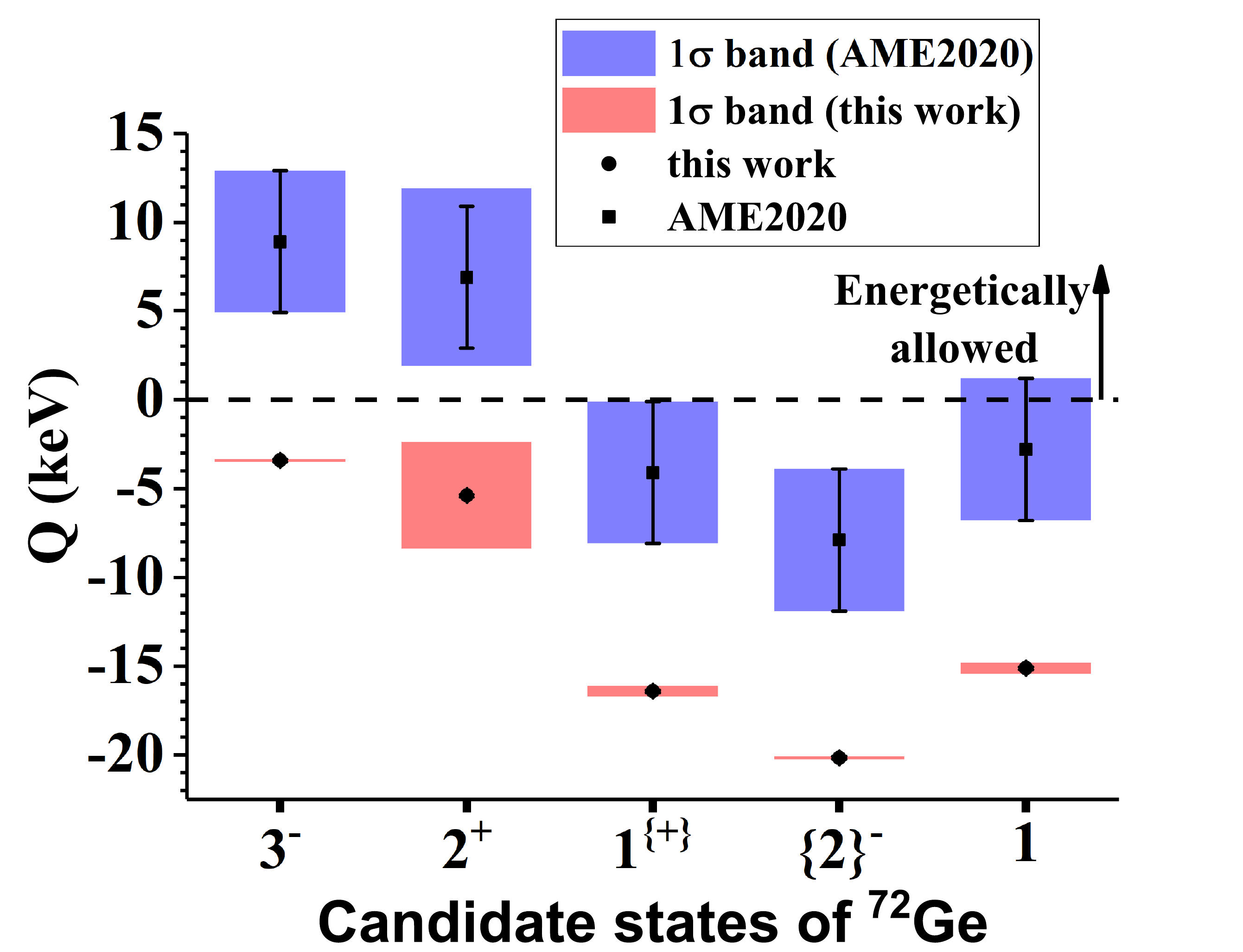}
   \caption{(Color online) {$Q$-values of the five potential candidate transitions of $^{72}$As ground state $\beta^+$/EC decay to the potential excited states in the daughter $^{72}$Ge from this work compared to the values derived using AME2020 $Q_{gs}$. The square and round points with 1$\sigma$ error bars use ground-state to ground-state $Q$-values from this work and AME2020, respectively, and show only the contribution from uncertainty of the ground-state to ground-state $Q$-value. The total uncertainty (including both the $Q_{gs}$-value and the excitation energy uncertainty) is indicated with shaded square areas.} }
   \label{fig:Q-value-comparison}
\end{figure}

Combining the new $Q_{gs}^{EC}$-value together with the nuclear energy level data gives the final $Q$-values for decays to the potential low $Q$-value states, see Table \ref{table:low-Q}. Also comparison to $Q$-values obtained with AME2020-values is given. The results are also plotted in Fig.~\ref{fig:Q-value-comparison}. 

The AME2020 $Q_{gs}^{EC}$-value, being 12.4~keV larger and having 4 keV uncertainty, could not unambiguously rule out these decays. Our results confirm that decays to any of the potential states are energetically forbidden. Decay to the 3327(3)~keV $2^+$ level is forbidden  at the $1.8\sigma$ level, and others more than $42\sigma$.

Though the EC/$\beta^+$-transitions studied here turned out to be energetically forbidden, this opens the door for the possibility to study another interesting decay type: radiative ``detour'' transitions. In addition to regular EC/$\beta^+$-decay, a second order process where a photon accompanies the lepton(s) is also possible. When the direct transition is hindered by angular momentum selection rules, a virtual transition via an excited state higher than the $Q$-value can contribute significantly, as pointed out in \cite{Longmire49}. Evidence of such a transition in $^{59}$Ni was seen in Ref.~\cite{Pfutzner2015}, where a transition via a state 26 keV higher than the $Q$-value was shown to contribute about 4\% of the experimental gamma spectrum. Since the probability of such a detour transition is proportional to $(E^*-E_{\gamma})^{-2}$ \cite{Longmire49}, where $E^*$ is the energy of the intermediate state and $E_{\gamma}$ the energy of the emitted gamma ray, a transition with a small negative $Q$-value would be optimal for the experimental study of detour transitions. In the case of $^{72}$As interesting transitions could proceed via the spin-1 state at 4358.7(3) keV. If this state turns out to have a negative parity, then the ground-state to ground-state transition could proceed via GT+E1 decay. However, finding transitions with even smaller negative $Q$-values and less competing transitions would be even better. Such transitions are likely to be found when other possible ultra-low $Q$-value transitions are investigated.

\section{Conclusion and Outlook}
A direct high-precision ground-state to ground-state  EC-decay $Q$-value measurement of $^{72}$As (2$^{-}$)$\rightarrow ^{72}$Ge (0$^{+}$) was performed using the PI-ICR technique at the JYFLTRAP Penning trap mass spectrometer. A $Q$ value of 4343.596(75)~keV  was obtained and its uncertainty was improved by a factor of 54.
A discrepancy of more than three standard deviation was found to the previously adopted value in the AME2020.
Our updated and significantly more precise $Q$-value is 12.4~keV smaller than the one adopted in AME2020.  
We confirmed that all five potential ultra-low Q-value decay transitions, one through EC and four through $\beta^+$ decay are energetically forbidden at least at the 1.8$\sigma$ level. This finding underlines the need to measure the $Q$-values to high precision before attempts to detect such possible low $Q$-value decay branches is made, with the goal to realize 
these decays for neutrino mass determination. While the negative $Q$-values exclude the use of $^{72}$As to study neutrino mass, the small negative $Q$-values could make it a candidate for the study of $\beta$-$\gamma$ detour transitions proceeding via virtual states.
%


\acknowledgments 
 We acknowledge the staff of the accelerator laboratory of University of Jyv\"askyl\"a (JYFL-ACCLAB) for providing stable online beam and J.~Jaatinen and R.~Sepp\"al\"a for preparing the production target. We thank the support by the Academy of Finland under the Finnish Centre of Excellence Programme 2012-2017 (Nuclear and Accelerator Based Physics Research at JYFL) and projects No. 306980, 312544, 275389, 284516, 295207, 314733 and 320062. The support by the EU Horizon 2020 research and innovation program under grant No. 771036 (ERC CoG MAIDEN) is acknowledged.
 

\bibliographystyle{apsrev4-1}

\bibliography{my-final-bib-from-jabref}
\end{document}